# Forming Teams for Teaching Programming based on Static Code Analysis

Davis Arosemena-Trejos[1], Sergio Crespo[2] and Clifton Clunie[3]

[1] Universidad Tecnológica de Panamá
Panama City, Panama
*davis.arosemena@utp.ac.pa*

[2] Universidade do Vale do Rio Dos Sinos
Sao Leopoldo, Brasil
*crespo.sergio@gmail.com*

[3] Universidad Tecnológica de Panamá
Panama City, Panama
*clifton.clunie@utp.ac.pa*

**Abstract**
The use of team for teaching programming can be effective in the classroom because it helps students to generate and acquire new knowledge in less time, but these groups to be formed without taking into account some respects, may cause an adverse effect on the teaching-learning process. This paper proposes a tool for the formation of team based on the semantics of source code (SOFORG). This semantics is based on metrics extracted from the preferences, styles and good programming practices. All this is achieved through a static analysis of code that each student develops. In this way, you will have a record of students with the information extracted; it evaluates the best formation of teams in a given course. The team's formations are based on programming styles, skills, pair programming or with leader.
**Keywords:** *Work team, Teaching programming, Programming styles, Static Code analyzer, Ontology.*

## 1. Introduction

The programming has been one of the areas of knowledge that demands more concentration, analysis, organized labor, patience and dedication, this is difficult for new students due to immaturity and other factors that cause specific problems in programming as is the case of the good implementation of languages, the correct use of the instructions, problems in the development of logic, creating optimal algorithms, among others [1]. On the other hand, there are other difficulties related to the social environment, heterogeneity and personal problems that affect the process of programming and therefore learning it [2].

To overcome these problems, efforts have been used for the development of methodologies and tools that support the teaching-learning process of programming, such as the research of [3] that present a framework for static analysis of programs for students, this returns, in case of errors, correction suggestions. [4] [5] present also tools to support the teacher in this process. There are techniques posed to improve the student's ability through good programming practices that allow them to develop skills in understanding programs [6]. On the other hand, [7] present a teaching method that Cognitive Apprenticeship Learning program helps students. The use of teams in classrooms is one of the most widely used methods that can actually help in this process [8] [9] [10]. However, these groups to be formed without taking into account certain considerations, may cause an adverse effect on the teaching-learning process of programming [1] [11].

This paper aims to present a tool for the formation of work teams based on static analysis of source code (SOFORG), which extracts information on programming styles, preferences and best practices through static analysis of the students develop programs, and likewise, if it existed, SOFORG provides an assessment of the problem suggested based on an ontology model assumption. The result of each analysis will be stored as static metrics for each student, in this way, there will be a complete record of the exercises applied in group or individual. For the formation of groups, the tool evaluates the current status of students in a particular course in order to suggest the best option of forming groups and these groups may be based on skills, programming styles, pair programming, groups with leader [12].

In addition to the above, SOFORG is useful in identifying gaps based on best practices of students [13] [14] (e.g. initialization of variables, variables that are not declared private or protected, others), which in most cases becomes a tedious task for the teacher, will also help keep a statistic on the percentage of students' progress that will assess the effectiveness of the techniques, methodologies, models and types of exercises used by the teacher groups, and other research within the teaching-learning programming process.

## 2. Static code analysis

Static code analysis is a technique used in Program Comprehension to program evaluation and analysis without being executed. This technique extracts relevant information about the program structure, programming styles, and other semantic errors [15]. Static code analysis has many applications in software engineering, but this is used as a technique within a few tools for teaching and learning of programming, such is the case of a framework for static analysis of programs for students [3],another tool assesses quality of programs through certain software engineering metrics [16], also, there are others who identify the algorithms used by students and assesses the ability of these [17], and likewise, [18] and [19] propose methods by static analysis for detecting errors and shortcomings. On the other hand, the static code analysis has a number of approaches that are presented by [17] these are: based on knowledge, similarity assessment, reverse engineering.

## 3. Team for teaching programming

Developing collaborative learning environments (work teams) through the achievement of mixed ability, it can be used as a teaching method for teaching programming [20]. In this way, students can be formed in teams to solve a programming assignment, where members discuss a possible solution and then they develop either by dividing sub-job work for each member or working together in a single module or task, if they can use pair programming on collaborative environment [8] or working on one computer [21]. Thus, this creates indirectly the effect that each student becomes the guardian of another [9], because they can consult different concerns with classmates or look together the work done by the rest. The use of teams can be more effective by instructional techniques, cognitive strategies or methodologies complete, however the success of the groups for teaching programming is not a trivial task, therefore sometimes it is not used very well, because there are internal and external factors that affect the performance and its success [2], [11].

## 4. Related research

[3] presents a framework for static analysis, where it is used to the student's practice, it allow to write better programs, because it gives assistance to the teacher in class and allows him to understand the real situation of students. This framework uses software engineering metrics and comparisons of models to assess the students and a program in case you find errors, notify the student and suggests a possible solution. On the other hand, [1] propose the formation of student's group for a collaborative learning programming, the formation of these groups is based on programming styles. In this analysis, it uses a tool called a Program Quality Assessment (PQA-C) that determines a percentage or value based on a set of metrics, where the highest scoring students form teams and the same way, the intermediate and lowest students form others groups.

## 5. Description of SOFORG

The SOFORG tool is aimed to assist the teacher in teaching programming. Its purpose is extract relevant information from students' source code through a static analysis on Java language and provides a record that allows it to create teams based on characteristics.

5.1 Extracted information

The result of extracted information is divided into 4 groups: programming styles, preferences structures, best practices and possible student outcome assessment, which will be described below.

5.1.1 Programing style

Programming styles are independent on the final functionality of the program, which these represent the appearance and format that each programmer gives to source code. There are many programming styles because they depend on the personality and habits programmer [2]. Some important programming styles with their respective metrics are present below based on [22], [23]:

**Length of identifiers (LI):** this takes into account names of variables, methods, classes, interfaces and packages, in which extracts an overall average of the length thereof. The following equation is proposed to be measured:

$$PL = \frac{NC}{NTN} \qquad (1)$$

where: *NC* - total number of characters of all identifiers, *NTN* - total number of identifiers found in the code *PL* - is the average length of identifiers.

**Indentation (I):** it is the space that exists at the start of each line of code; this is used to improve eyesight and reading it. The following equation is proposed for measurement:

$$PI = \frac{\sum_{l=1}^{L}\sum_{i=1}^{I}(NE_i)}{\sum_{l=1}^{L}(N_l \cdot 4)} \quad (2)$$

where: *L* - is the total number of lines of code, *I* - is the number of spaces of indentation for each line of code *l*, $NE_i$ - is the number of space indentation of the source code, $N_l$ . 4 - is the nesting level of each line of code (0,1 ..) multiplied by 4, *PI* - is the ratio between $NE_i$ and $N_l$ . 4. The latter sum is that is referenced to a 4-space indentation, for every level of nesting, e.g. line: 2 on level: 0 - it has 0 space indentation, line: 14 on level: 2 - it has 8 spaces indentation. This indentation style is presented in some Java IDEs like as Netbeans.

**Use curly bracket (CB):** this indicates that a curly bracket (usually to state structures, classes, methods) may be in the same line as the structure is declared or in the next line.

**Use blank lines (BL):** it represents the blank space between lines of code; this gives a better view and formats it. Many students vary the amount of blank lines, especially for separating structures or lines of code in a section. The metric is proposed for this style below:

$$PB = \frac{\sum_{b=1}^{B}(LB_b)}{\sum_{l=1}^{L}(L_l)} \quad (3)$$

where: *B* - is the possible repetitions of blank lines in the code, $LB_b$ - total number of blank lines, $L_l$ - is the total number of lines that have the code, *PB* - is the ratio between the total number of blank lines and the total of lines that have the code.

**Statement of code per line (SCL):** it describes the amount of code lines that is completed by semicolon (``;"), also this can represent either a part or all of a structure that can be in one line. The following metric for the measure is proposed below:

$$PC = \frac{\sum_{l=1}^{L}(L_l)}{\sum_{c=1}^{C}(LR_c)} \quad (4)$$

where: *C* - is the possible number of reference code lines, $LR_c$ - is the *c* times existing reference lines in the code, *PC* - is the ratio between the total code lines and reference code lines.

There can be three types of reference code lines and they are based on the default format of some Java IDEs like Netbeans: 1. sentences ending by semicolon, 2. declarations of structures, methods or classes followed by the curly bracket, 3. closing curly bracket. Example: *if (var < 2) {return var; } ...* has 3 reference code lines.

**Documentation program (D):** basic programming students do not have a habit of documenting their code, because they find it boring or unnecessary use. The next formula is proposed for measuring of this programming style:

$$PD = \frac{NCD}{\sum_{s=1}^{S}(LS_s)} \quad (5)$$

where: *NCD* - is the total number of characters within the code documentation, *S* - number of code lines without computing the documentation lines, $LS_s$ - is the total number of lines undocumented code represented by the *s* possible times.

**Initialization variables (IV):** it may be optional, but in some languages such as *C*, initialize all variables when their declaration is of vital importance in the elimination of garbage rows of memory, but this problem does not exist in the new programming language. This programming style is measure calculating the total percentage of initialized variables.

### 5.1.2 Preferences of structure

Programming preferences are classified as programming style, but to this work, has been made a separate classification to a better appreciation. There are multiple ways to solve a problem within the program, all can be equally efficient. These multiple ways can be: decision and repetition structures, recursive or inductive methods and specific functions of language. The use of one of them is the programmer's preference.

This tool only evaluates preferences in decision and repetition structures, these are 4 types: *Ifelse_elseif (IE), Ifelse_elseif_switchcase (IES), For_while_dowhile (FWD), While_dowhile (WD)*.

### 5.1.3 Best practice

Similarly, SOFORG evaluates the source code to suggest best practices. In programming, best practices represents the set of patterns or styles of programming that the student must apply to improve performance and maintainability of the programs, and avoid programming errors.

This information is useful to identify potential deficiencies, especially in OOP, from the evaluation of those best practices that the student has not applied. The analysis of best practices is knowledge-based approach, as is the preference structures, explained in the section 2. The following are best practices that SOFORG can evaluate:

**Attributes that have not been initialized in the constructor:** It is best practice of OOP, where the attributes have to be initialized in the constructor of their respective class, except *static* or *final* attributes and that belong to abstract classes or interfaces.

**Public static type attributes uninitialized in his statement:** All attributes of *public static* type must be initialized in its declaration, no matter they are initialized in a constructor, because this could produce programming error when used by another class and has not been initialized.

**Final type attributes must be declared static:** When an attribute is declared *final*, but not static, will cause each instance of the class to which the attribute belongs, keep a record in memory of the same value, but if the attribute is declared *static*, there will be a single record for all instances. Thus, this will help in better utilization of memory.

**Attributes that are not declared private or protected:** All attributes except those declared *final* or *static* and the attributes of the interfaces must be declared *private* or *protected*.

**Class without constructor:** Each created class should have its own constructor for purposes of maintainability, except the classes with main method, abstract classes and interfaces.

**Methods can be declared static:** The use of static methods improve the performance of compilers and consume less memory, since no copies are created for each instance method.

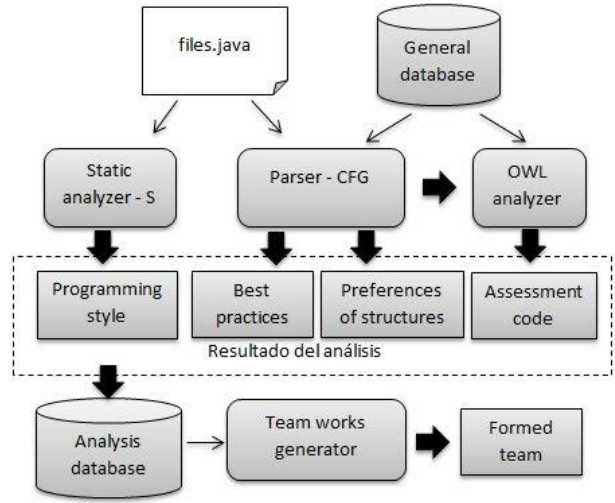

Figure 1: Outline of SOFORG elements.

**Nested classes can be declared static:** he use of static classes are similar to the use of the methods of this type, thus this represents a best practice.

### 5.1.4 Assessment code

SOFORG used similarity assessment approach (described in section 2) to the analysis of assessment code, because the students' program is compared with a stored model for it. These models contain the possible solutions of an exercise, this is an ontological structure written in OWL [24]. The exercises will be graded based on 100 %, so that each part of the structure will have a weight of the problem, for example, if an model implements a cycle, it will has a weight of 13 % of the total 100, the operations within this cycle will have another percentage and so on. If the logical statement of cycle is erroneous, also the nested operations and so this is a total loss of all points in the cycle and operations, although they will be well. On the other hand, the position and sequence of structures and operations within the program will have others weight.

Not all problems that are presented to students could have stored models in OWL. This depends on the complexity, use of special features of Java and GUI. Therefore, in case there is no model, the teacher evaluates and enters the assessment where it will be stored with full registration of analysis.

### 5.2 Static code analysis for students

SOFORG has some elements that form the tools, these elements are showed in the figure 1. The *Static analyzer - S* element is responsible for extracting the programming style metrics, it scans the source code in form string. On the other hand, *Parser – CFG* analyzes with Context-free grammar using a parser and interpreter called *JavaCC* [25], this element provides the result of good practices and preferences of structures, also if there is a solution model of exercise, it grades the student's program and provides an assessment. For this assessment, *Parser - CFG* extracts the program components of student model and sends them to *OWL analyzer*. The *OWL analyzer* implements *Jena* [26], which is a framework for creating web semantic application where this receives components program information from *Parser-GLC* and generates OWL instances of these, then it loads the solution models of exercise by reference that is specified in the source code.

The solution models of exercise are ontological structures that represent the possible solutions of these exercises, this ontological structure is written in OWL and they are represented by an OWL file. The benefits of OWL are to organize solutions through classes and subclasses, for example: Super class: Solution1-2 has subclasses: Solution1 and Solution2. Although all applications of this type are NP order, the organizing of solution in this way improves the search time. Therefore it locates the right solution into ontological structure then compares the properties of the found components (OWL generated instances) with the components of the respective solution for grading of this, all through an ontological reasoner in *Jena*.

The following lists the components that can be identified with their respective properties in OWL:

*Inputs:* logical order (before, after), the variable that receives the input, the hierarchy level and its location (if it is nested within a structure of if-this, while and other, but in the case of being if-else, if it is in the IF or ELSE).

*Prints:* logical order, variables that are printed, hierarchical level and location.

*Operations:* logical order, used variables, used operators, the variable that was assigned the result, hierarchical level and location.

*Numeric and Boolean values:* it does not contains properties, it use his name for this.

*Logical statements:* used variables, logical operators, if they relate to another statement through AND or OR.

*Variables:* type of variable, initialization, which components it was used.

*If-else structure:* logical order, used local statements, hierarchical level and location.

*While structure:* logical order, used local statements, hierarchical level and location.

*Do-while structure:* logical order, sentence used, hierarchical level and location.

*Switch-case structure:* logical order, reference variable, values of the case, hierarchical level and location.

*Array:* logical order, type, length, hierarchical level and location.

*Methods:* variable types that accept, whether or not return a result, result type that return, location.

*Class:* class type (abstract, public, private), whether inherited or not, whether it implemented or not a interface.

*Attributes:* privacy type, variable type, initialization, components where it was used.

*Constructors:* variable types that accept, source class, type of privacy.

*Instances:* location, source class, constructor to access.

5.3 Generation of work teams

This tool has a set of options that the teacher may choose to form groups and their possible characteristics, based on [21], [1], [27], some types of team with application benefits are presented as follows:

**Pair programming teams:** this type of team uses the same computer, includes two students sitting in front of it, where one takes the role of *Driver* and the other the *Navigator* role; both can discuss a possible solution, but the *Driver* is the one responsible for writing the code in the computer and the *Navigator* reviews and monitors the code already written for finding errors. The optimal use will depend of these students possess the same programming styles and skills.

**Teams based on leading student:** this is comprised of students where one or more have the role of leader, it may be advantageous to use because the leading students provide better support to their classmates. Students receive from their leader a clarification of their doubts and more focused and personalized explanations than the teacher. They are useful in those classrooms where most students

```
if (porcenEstilo >= 75)
        //FORMATION PAIR TEAMS BASED
        ON PROGRAMMING STYLE
else if (porcenCapaDife > 45 and porcenCapaDife < 55)
        //FORMATION PAIR TEAMS BASED
        ON DIFFERENT ABILITIES
else if (porcenCapaIgual > 75)
        if (numEstuCapaTutor >= numerosDeGrupos)
                //FORMATION TEAMS OF 5 BASED
                ON SIMILAR ABILITIES WITH LEADER
        else
                //FORMATION TEAMS OF 5 BASED
                ON SIMILAR ABILITIES WITHOUT
                LEADER
else if (numEstuEstiTutor >= numerosDeGrupos)
        //FORMATION TEAMS OF 5 BASED
        ON PROGRAMMING STYLE WITH LEADER
else
        //FORMATION TEAMS OF 5 BASED
        ON PROGRAMMING STYLE WITHOUT
        LEADER
```

Figure 2: Algorithm for the suggestion of the best option forming group.

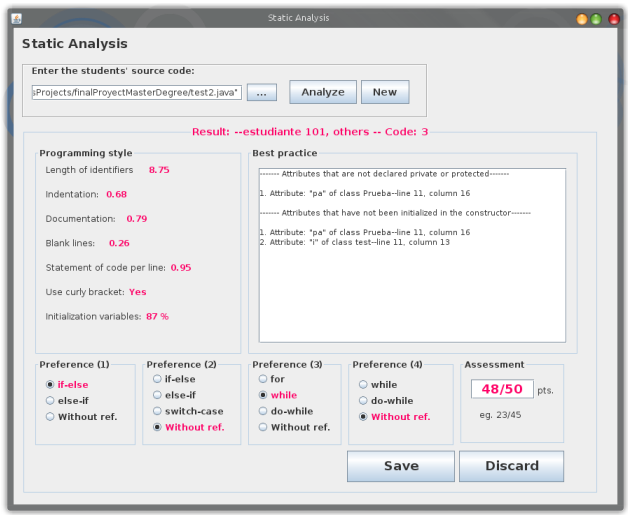

Figure 3: View of static analysis result.

have little or lower capacities compared to the rest of the group.

**Teams based on programming styles:** it is comprised of students who have the same programming style. This type of group will enable better integration and adaptation in the tasks assigned to students sharing the same preferences and styles of programming.

**Teams based on abilities:** these teams are useful for grouping students with similar or different abilities. Grouping students with similar abilities may help to increase the performance and production in the programming work, but on the other hand, teams that are grouped by different abilities may help to students with lower ability in the learning process.

**Random teams:** these teams are composed of students who have been randomly generated, regardless of any criteria. Its main advantage is that students are fully distributed in such manner that helps them to develop the skills needed to work in groups.

Professor not always know what is the best type of group to form, this is because it depend on the current status of students in a course, and it can be ability levels and styles of programming [12]. The ability of students is registered in each static analysis; this is based on the evaluation provided and a weight that is given to the suggestions of best practice identified in solution students. The metric of ability will measure with values between 0 and 100. Thus, SOFORG provides the option to choose the best team to form; this is made by evaluating the current status of the course. An algorithm for the suggestion of the best option forming group is presented on the figure 2, this represents the structure of decision that is based to suggest that team formation, where: *porcenEstilo* - is the percentage of similarity of programming styles in students of the course, *porcenCapaDife* - is the percentage of students with abilities larger than average capacity of the course, *porcenCapaIgual* - percentage of students with similar abilities, *numEstuCapaTutor* - is the number of students with abilities greater than average range, *numEstuEstiTutor* - is the number of students with abilities greater than 91. The values, ranges and percentages used in the algorithm can be edited in SOFORG, this allows the tool to raise awareness in verifying the similarity of programming styles, abilities and detecting leading students, and likewise, there are other variables that can be edited such as: the size of team to form that has by default 5 members.

This algorithm evaluates the pair teams formation and then the formation teams of five with either tutor or not, because formations with pair teams and based on programming style get more benefit for students in the teaching-learning process [1], [5].

5.4 Features of SOFORG

SOFORG is a desktop tool with database on a central server; each teacher will start its own session. When logged on, the system queries could be done through existing forms or start a new static analysis specifying the java files. Once these files is specified, it analyzes showing the result, this result can be changed as shown in Figure 3, this figure shows the result of analysis that is organized in their respective categories (detailed in sections 5.1), together with the student's name and code of exercise. Also the programming styles panel presents the results based on the metric set, the preferences are separated in each panel and explained, and use radio buttons for each option, in case there is no particular preference, SOFORG marks the radio button "Without ref.", on the other hand, the panel of best practice is a text box that shows the best practice that students have not applied specifying, depending on the case, the referenced object, class to which it belongs, line and column in the code, and finally, the assessment panel which has a text field and have the following format *obtainedPoints*/*totalPoints*, if the assessment does not appear in the text field, the teacher should include his assessment for student exercise. When many analyses are stored, the teacher can generate teams using these tools, the figure 4 shows 4 panels for the setting of forming teams. The course code should be included as shown in the 1 panel, then the teacher can choose several options for forming teams such as: type of team, number members, if the teams is based on abilities, programming style or randomly, when the options has been set, the tool shows the selected options as shown in the 4 panel. The teams are generated and it presents the name and ID of students within the team to which they belong.

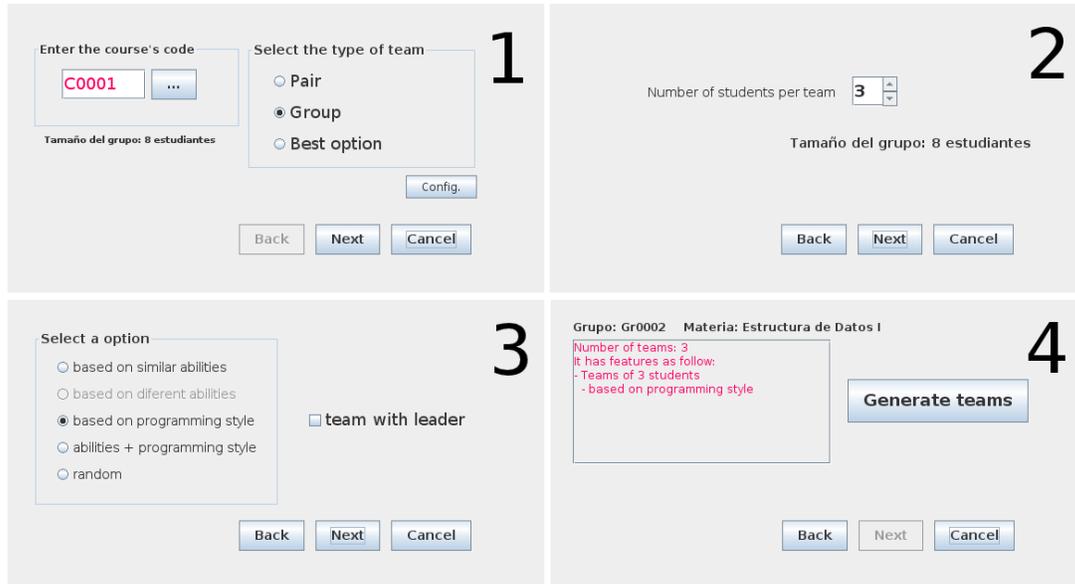

Figure 4: View of the setting for the formation of teams.

5.5 Other contributions of SOFORG

SOFORG can identify applied exercises and hold an annual cycle of them, also permits viewing the progress of individual students by grade, level, from the student evaluation, and likewise, this tool helps in identifying group and individual deficiencies associated with the detection of best practice. Also it allows evaluating the effectiveness of teaching methods used by teachers, from students' progress. Also, know the problems more difficult for students, in this way; the teachers can evaluate the use of the best problems to start a class. All this will help to hold a complete history that allows for other scientific studies.

6. Case study

A case study to evaluate the SOFORG tool was carried out in the second semester of 2011 with one group of fourteen students and a course of basic Java language at Technological University of Panama. Two exercise of programming was applied to students individually, and then source codes of the students were analyzed with SOFORG for forming the teams finally. The setting for forming teams is three students per teams and based on programming styles. The list of this students can be seen in the table 1 where this has the metrics of programming styles and preferences as result of static analysis of both exercises.

The figure 5 shows the result of formed teams by SOFORG and also the setting final in red letters.

4. Conclusions

A tool to the formation of work teams based on static analysis of source code (SOFORG) was proposed, together with the description of its most important elements and characteristics, two types of static analysis was used, these implemented context-free grammar and pattern recognition.

Table 1: List of students with their metrics from static analysis of SOFORG

| No. | Students name | LI | I | CB | BL | SCL | D | IV | IE | IES | FWD | WD |
|---|---|---|---|---|---|---|---|---|---|---|---|---|
| 1 | Davis Nelson | 5.82 | 0.03 | 0 | 0.29 | 1.20 | 0.51 | 66 | ifelse | elseif | for | dowhile |
| 2 | Francisco Lau | 6.41 | 0.01 | 0 | 0.11 | 1.26 | 1.60 | 46 | elseif | switch | while | null |
| 3 | Jairo Ramos | 5.76 | 0.02 | 0 | 0.05 | 1.17 | 0.53 | 50 | ifelse | elseif | for | dowhile |
| 4 | Jesus Samaniego | 4.13 | 0.02 | 0 | 0.07 | 1.10 | 1.57 | 76 | ifelse | elseif | while | null |
| 5 | Nidia Salceda | 4.33 | 0.21 | 0 | 0.40 | 1.28 | 0.46 | 33 | ifelse | switch | dowhile | null |
| 6 | Steven Caballero | 5.88 | 0.06 | 0 | 0.01 | 1.18 | 0.95 | 75 | elseif | null | for | dowhile |
| 7 | Eduardo Griffith | 5.28 | 0.20 | 1 | 0.13 | 1.10 | 0.45 | 73 | elseif | elseif | while | null |
| 8 | Francisco Him | 5.17 | 0.01 | 0 | 0.04 | 1.26 | 1.77 | 30 | ifelse | null | while | while |
| 9 | Fidelino Camarena | 6.36 | 0.01 | 0 | 0.25 | 1.30 | 1.68 | 43 | elseif | elseif | dowhile | dowhile |
| 10 | Franklin Rodriguez | 5.75 | 0.97 | 0 | 0.28 | 0.97 | 3.14 | 36 | null | null | null | null |
| 11 | Frederick Sanson | 4.62 | 0.0 | 1 | 0.17 | 1.30 | 1.44 | 57 | elseif | elseif | dowhile | dowhile |
| 12 | Israel Gamas | 5.51 | 0.23 | 0 | 0.08 | 1.18 | 0.34 | 75 | elseif | elseif | while | while |
| 13 | Juan Munoz | 4.45 | 0.05 | 0 | 0.21 | 1.36 | 0.43 | 47 | elseif | elseif | dowhile | dowhile |
| 14 | Yeiluant Flores | 8.64 | 0.14 | 0 | 0.44 | 1.28 | 8.91 | 34 | elseif | elseif | dowhile | null |

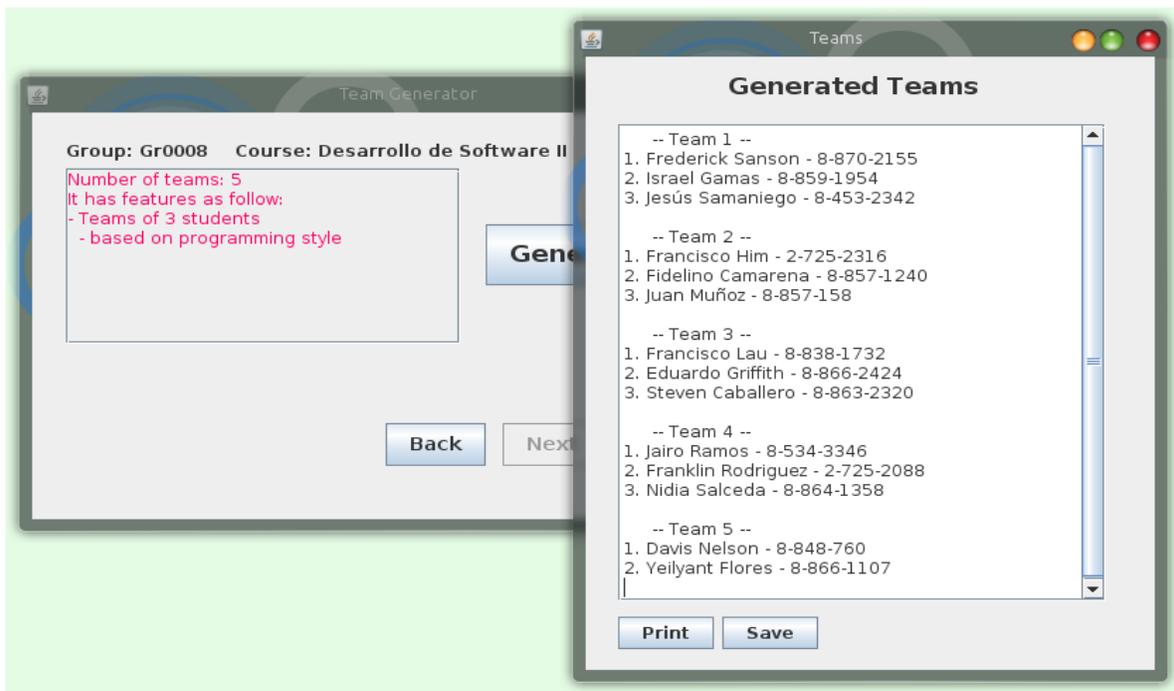

Figure 5: Generated teams view.

Also, there are ontological models for the evaluation of the problems, but not all problems will have an ontological model, it depends on the complexity and the several ways of algorithms that can be employed.

This tool can be implemented in the universities to submit basic programming courses, adding other modules to complement the labor of teachers, on the other hand, SOFORG is flexible to work with more parameters (programming styles, preferences) to allow capture of more precisely the characteristics of students, and likewise, you can add other rules in the detection of best practices and program models to the repository.

The better your paper looks, the better the Journal looks. Thanks for your cooperation and contribution.

## Acknowledgments

Work funded by Secretaria Nacional de Ciencias y Tecnología (SENACYT) of Panama through the Proposal with No. APY-GC10-026B.

**Davis Arosemena-Trejos** is a researcher at the Technological University of Panama. He has a Master's degree (MSc) in Information and Communication Technology from the Technological University of Panama. His research interests include Software Engineering, Semantic Web, programming's teaching-learning process and other.

**Sérgio Crespo** is a professor at the Universidade do Vale do Rio Dos Sinos – Brasil, PhD awarded by Pontifícia Universidade Católica do Rio de Janeiro.

**Clifton Clunie** is a professor at the Technological University of Panama, PhD awarded by the Universidade Federal do Rio de Janeiro.